\documentclass[pra,aps,superscriptaddress,nofootinbib,twocolumn]{revtex4-1}

\usepackage{mathrsfs}
\usepackage{amsfonts}
\usepackage{amssymb}
\usepackage{amsmath}
\usepackage{graphicx}
\usepackage[usenames,dvipsnames]{color}
\usepackage[colorlinks=true,citecolor=blue,linkcolor=magenta]{hyperref}
\usepackage{ulem}
\usepackage{lmodern}
\usepackage{amsthm}

\newcommand{\figpath}{.}

\newcommand{\ket}[1]{\vert{ #1 }\rangle}

\begin{document}

\title{Fault-tolerant fidelity based on few-qubit codes: Parity-check circuits for biased error channels}

\author{Dawei Jiao}
\affiliation{Graduate School of China Academy of Engineering Physics, Beijing 100193, China}

\author{Ying Li}
\email{yli@gscaep.ac.cn}
\affiliation{Graduate School of China Academy of Engineering Physics, Beijing 100193, China}

\begin{abstract}
In the shallow sub-threshold regime, fault-tolerant quantum computation requires a tremendous amount of qubits. In this paper, we study the error correction in the deep sub-threshold regime. We estimate the physical error rate for achieving the logical error rates of $10^{-6} - 10^{-15}$ using few-qubit codes, i.e.~short repetition codes, small surface codes and the Steane code. Error correction circuits that are efficient for biased error channels are identified. Using the Steane code, when error channels are biased with a ratio of $10^{-3}$, the logical error rate of $10^{-15}$ can be achieved with the physical error rate of $10^{-5}$, which is much higher than the physical error rate of $10^{-9}$ for depolarising errors. 
\end{abstract}

\maketitle

\section{Introduction}

Quantum computation has the capability to solve problems that are intractable in the conventional paradigm. The unique properties of quantum computation allow us to find quantum algorithms that are superior to classical ones, such as factorisation, search and quantum simulation algorithms~\cite{Nielsen2010}. Many quantum computation applications rely on the accurate manipulation of highly-entangled multi-qubit quantum states. For instance, solving the factorisation problem on the code-breaking scale requires an error rate of $\lesssim 10^{-12}$ per logical quantum gate~\cite{Fowler2012, OGorman2017}. Quantum error correction is a promising approach to the high-fidelity quantum computation. Taking the surface code as an example, by encoding the logical information in a two-dimensional array of qubits, the probability of a logical fault decreases exponentially with the array size when the physical error rate is lower than the threshold of $\sim 1\%$~\cite{Fowler2009, Wang2011}. Qubit initialisation, measurement, single-qubit and two-qubit quantum gates with sub-threshold error rates have been demonstrated with superconducting qubits and trapped ions~\cite{Barends2014, Ballance2016, Gaebler2016}. Given a quantum processor with sufficient qubits operated in the sub-threshold regime, we can implement any quantum algorithm with an adequately high fidelity. 

\begin{figure}[tbp]
\centering
\includegraphics[width=1\linewidth]{\figpath 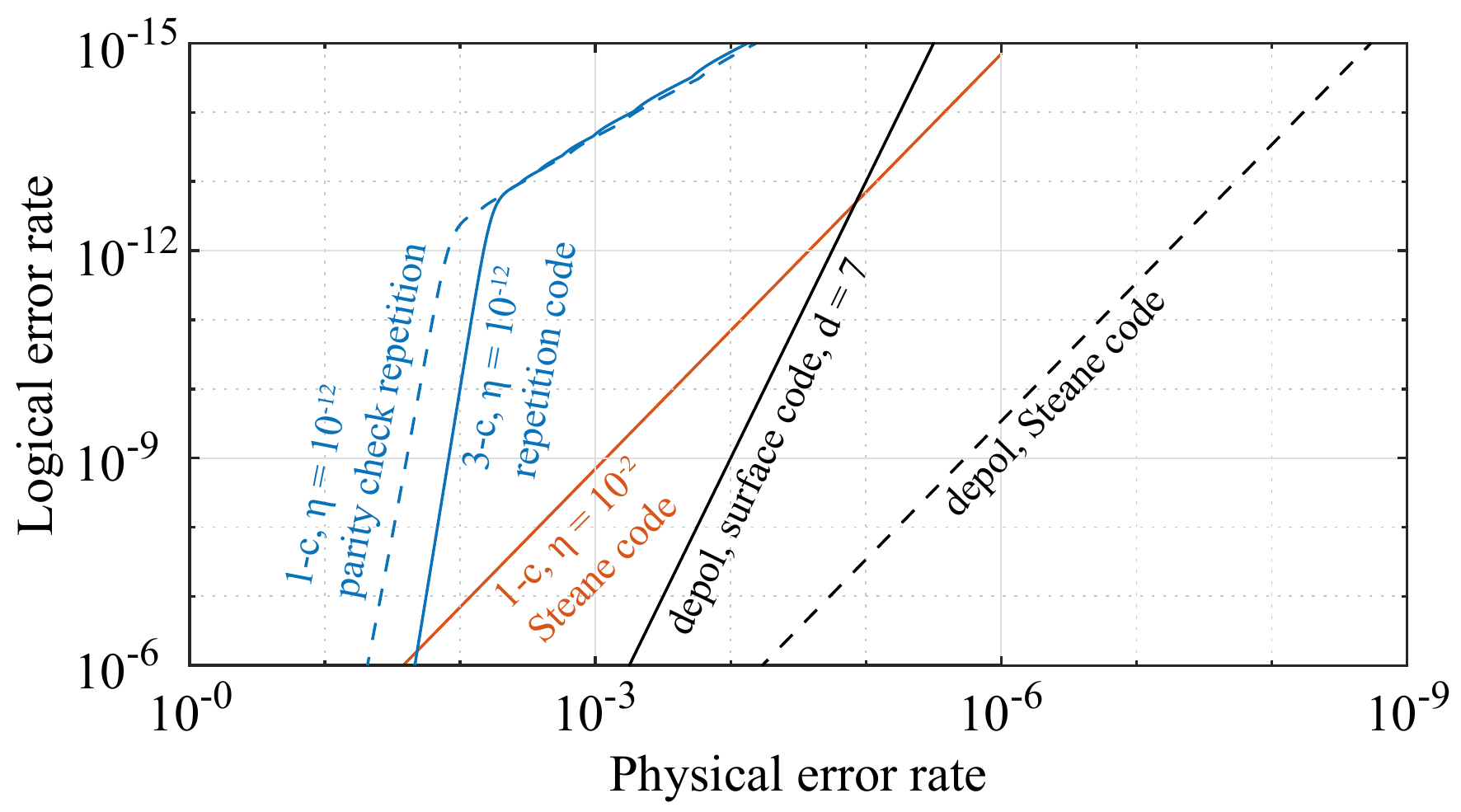}
\caption{
The logical error rate for varies codes and error models. In the parity check repetition, the encoding is not used, and errors are detected by repeating the two-qubit parity check measurements. The code distance of the repetition code is smaller than $22$. For the error model, depol, 1-c and 3-c represent the depolarising error model, the 1-channel dominant model and 3-channel dominant model, respectively. $\eta = \epsilon/p$, is the bias ratio between the error rate of dominant errors $p$ and the error rate of other errors $\epsilon$. 
}
\label{fig:summary}
\end{figure}

Quantum error correction is costly in the shallow sub-threshold regime. When the physical error rate is not adequately lower than the threshold, we need thousands of physical qubits for encoding one surface-code logical qubit, in order to achieve the logical error rate of $10^{-12} - 10^{-15}$~\cite{Fowler2013}. In recent years, rapid progress has been made in experiments. The qubit number and gate fidelity have been greatly improved~\cite{Google}, which promote the research of practical applications on intermediate-scale quantum computers in near term. Networked architectures can help to scale up the fault-tolerant quantum computer by connecting small processors using entanglement generation and distillation in addition to varies codes and error models~\cite{Li2012, Nickerson2013, Monroe2014}. Besides the increment in the qubit number, the gate fidelity in various quantum computation platforms has been constantly improved in the past twenty years~\cite{link}. By exploring the physics of non-Abelian anyons in materials, theoretical studies suggest that the error rate in topological quantum computation could be much lower than in conventional approaches~\cite{Nayak2008}. The development of technologies that provide lower physical error rates will reduce the encoding cost in the fault-tolerant quantum computation. In this paper, we study the quantum error correction in the deep sub-threshold regime, i.e.~in which the error rate is much lower than the threshold such that the fault-tolerant-level logical error rate can be achieved with encoding in only a few qubits. 

In this paper, we estimate the physical error rate required for achieving the fault-tolerant-level logical error rate using few-qubit codes. We consider the surface codes with small code distances and the Steane code for depolarising errors, and the Steane code and short repetition codes for biased Pauli-error channels. Usually, the error correction is more efficient for biased errors compared with depolarising errors~\cite{Aliferis2008,Brooks2013,Tuckett2018,Xu2019,Tuckett2020}. For biased channels, we focus on two cases: one or three of $15$ Pauli-error channels are dominant. For the 1-channel case, we find an error-correction circuit for each of the 15 Pauli-error channels, such that the dominant errors result in only one species of errors, e.g.~measurement errors. The correction of one species of errors is more efficient than general errors. Efficient circuits for the 3-error case are also identified. Using these circuits in the repetition code, the fault-tolerant-level logical error rate can be achieved with a relatively high physical error rate, but error channels need to be extremely biased. The Steane code can correct general errors. We find that error correction circuits of the Steane code are also efficient for correcting specific one-channel errors. The numerical results show that the physical error rate required by the Steane code is hundreds of times higher when the bias is of a ratio $10^{-2}$ than the rate when errors are depolarising. All results are summarised and shown in Fig.~\ref{fig:summary}. 

This paper is organised as follows. In Sec.~\ref{sec:error_model}, the Pauli error model is introduced. In Sec.~\ref{sec:circuits}, we discuss circuits that are efficient for different biased error models. In Sec.~\ref{sec:error_rate}, the definition of logical error rate used in this paper is given. The error correction for the depolarising error model using small surface codes and the Steane code is discussed in Sec.~\ref{sec:depol}. The error correction for 1-channel dominant models and 3-channel dominant models are discussed in Secs.~\ref{sec:1c} and \ref{sec:3c}. A summary of results is given in Sec.~\ref{sec:summary}. 

\section{Error model}
\label{sec:error_model}

In this paper, we focus on the deep sub-threshold regime, in which physical error rates are much lower than the threshold. In all the computation operations, we are interested in the case that the error rate of two-qubit gates is much higher than single-qubit operations, i.e.~state preparation, preparation and single-qubit gates. If errors caused by single-qubit gates are negligible, we can use the Pauli twirling~\cite{Bennett1996,Emerson2007,Dankert2009,Geller2013} to convert errors in two-qubit Clifford gates, e.g.~controlled-NOT and controlled-phase gates, into Pauli errors. Therefore, we model errors as follows: Single-qubit operations are all error-free; we only use the controlled-NOT gate in the quantum error correction, and the controlled-NOT gate with error is modeled as an error-free gate followed by the erroneous operation 
\begin{eqnarray}
\mathcal{N} = \sum_{\sigma_{\rm c},\sigma_{\rm t}} P_{\sigma_{\rm c},\sigma_{\rm t}} [\sigma_{\rm c}\otimes\sigma_{\rm t}],
\end{eqnarray}
where $\sigma = I,X,Y,Z$ are Pauli operators, ${\rm c}$ and ${\rm t}$ respectively denote the control and target qubits, $P_{\sigma_{\rm c},\sigma_{\rm t}}$ is the rate of the Pauli channel $[\sigma_{\rm c}\otimes\sigma_{\rm t}]$, and $\sum_{\sigma_{\rm c},\sigma_{\rm t}} P_{\sigma_{\rm c},\sigma_{\rm t}} = 1$. Here, $[U](\bullet) = U \bullet U^\dag$. Except the channel $[I_{\rm c}\otimes I_{\rm t}]$, the other 15 channels cause Pauli errors. $P_{I_{\rm c},I_{\rm t}}$ is the gate fidelity, and $1 - P_{I_{\rm c},I_{\rm t}}$ is the total error rate. For the depolarising error, 15 Pauli errors are uniformly distributed, i.e.~rates of all Pauli errors are the same, and $P_{\sigma_{\rm c},\sigma_{\rm t}} = (1 - P_{I_{\rm c},I_{\rm t}})/15$, where $\sigma_{\rm c}\otimes\sigma_{\rm t}\neq I_{\rm c}\otimes I_{\rm t}$. For the extremely biased error, we consider two cases. In the 1-channel case, only one of 15 Pauli error channels is dominant. In the 3-channel case, three of 15 Pauli error channels are dominant. 

\section{Parity-check circuits}
\label{sec:circuits}

\begin{figure}[tbp]
\centering
\includegraphics[width=1\linewidth]{\figpath 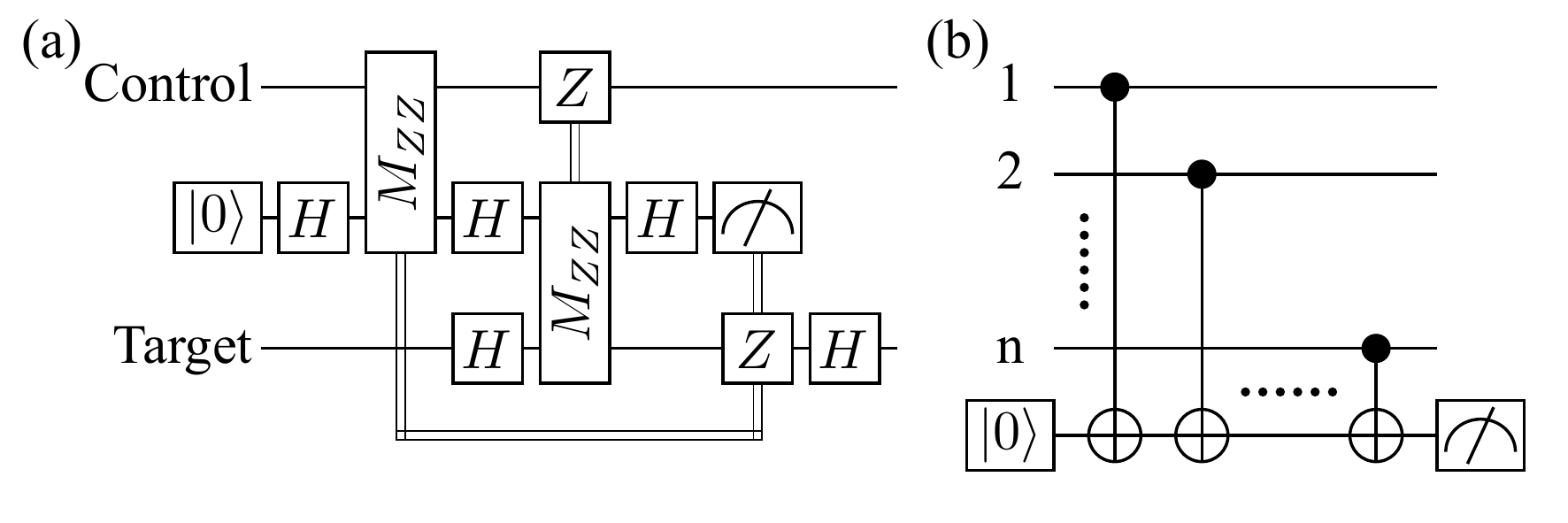}
\caption{
(a) Controlled-NOT gate realised using two-qubit parity checks. $M_{ZZ}$ is the projective measurement of $Z\otimes Z$. The double lines denote feedback gates. (b) A parity-check circuit for measuring $Z\otimes Z\otimes \cdots \otimes Z$ of $n$ qubits based on controlled-NOT gates. 
}
\label{fig:circuit_CandP}
\end{figure}

In the quantum error correction of a stabiliser code, errors are detected by measuring a set of generators of the stabiliser group, which is a subset of Pauli operators. For example, in the Steane code, the stabiliser generators are four qubit Pauli operators in the form $X\otimes X\otimes X\otimes X$ and $Z\otimes Z\otimes Z\otimes Z$. To implement the fault-tolerant quantum computation using a stabiliser code, we must find proper parity-check circuits for measuring these stabiliser generators~\cite{Gottesman1998}: Errors generated in the circuits must be detectable and correctable by the measurement outcomes, and these errors must be prevented from spreading in the circuit, which could transform a single-qubit error into a two-qubit error. The measurement of a generator also needs to be a projective measurement, i.e.~if the measurement outcome of the generator $g$ is $\mu = \pm 1$, the state $\rho$ is transformed to $[(\openone+\mu g)/2]\rho$ up to a normalisation factor. We note that when the circuit is implemented with errors, the measurement is not always exactly projective. 

\begin{figure}[tbp]
\centering
\includegraphics[width=1\linewidth]{\figpath 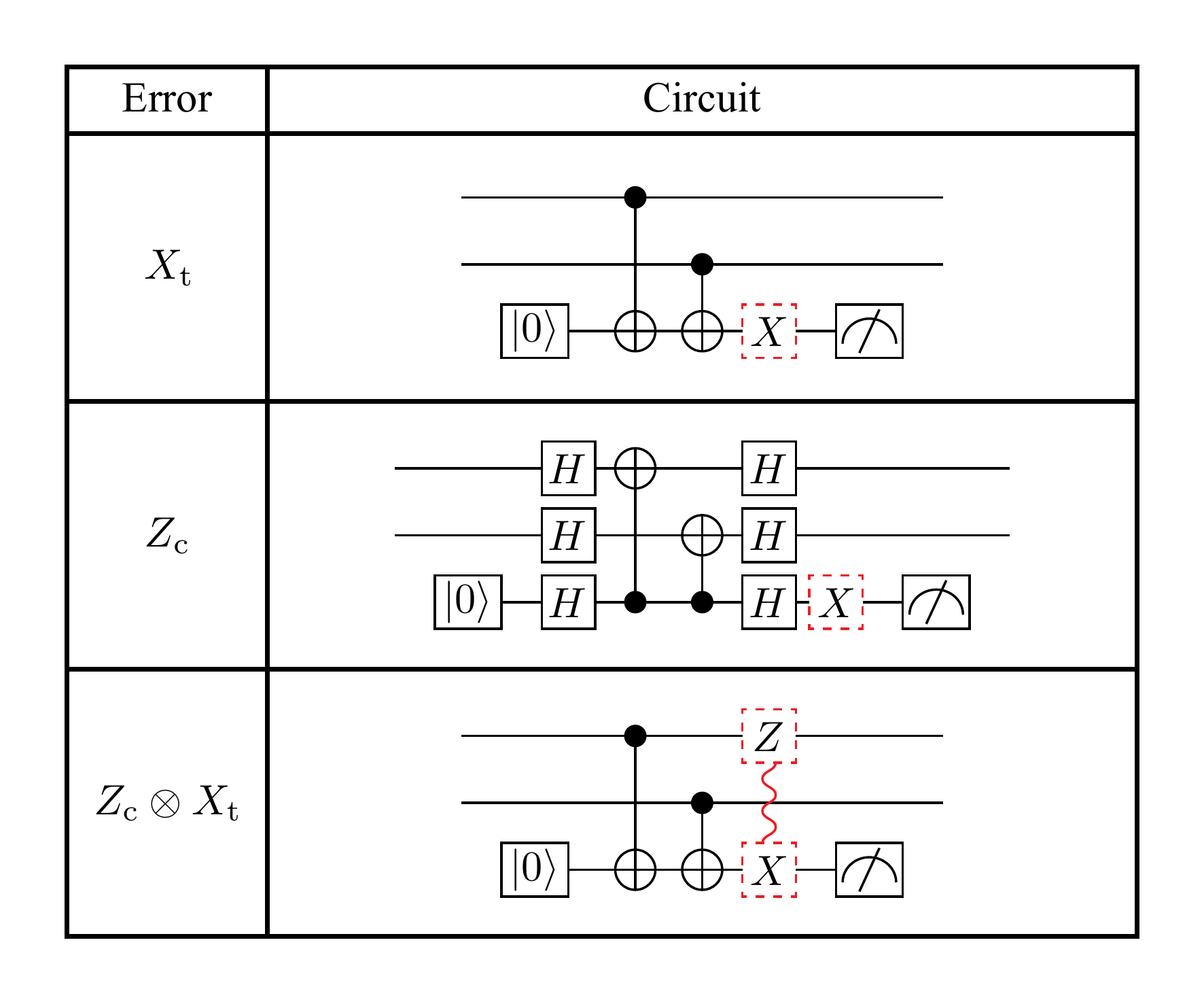}
\caption{
The 1-channel efficient circuits for $X_{\rm t}$, $Z_{\rm c}$ and $Z_{\rm c}\otimes X_{\rm t}$. The dashed red square denotes the error that may happen. Two errors connected by the wavy line always occur simultaneously. 
}
\label{fig:circuit_1cA}
\end{figure}

In this section, we propose parity-check circuits tackling biased error channels, i.e.~the 1-channel and 3-channel cases. We consider the measurement of two operators $Z\otimes Z$ and $Z\otimes Z\otimes Z\otimes Z$. The measurement of other two-qubit and four-qubit Pauli operators can be obtained by modifying circuits for measuring $Z\otimes Z$ and $Z\otimes Z\otimes Z\otimes Z$, by inserting single-qubit Clifford gates before and after the measurement to adapt the Pauli basis. With an ancillary qubit, we can implement a controlled-NOT gate using two projective measurements of $Z\otimes Z$ and single-qubit operators [see Fig.~\ref{fig:circuit_CandP}(a)]. Therefore, replacing the controlled-NOT gate with the $Z\otimes Z$ measurement, the $Z\otimes Z$ measurement and single-qubit operations form a universal gate set for the quantum computation. 

\begin{figure}[tbp]
\centering
\includegraphics[width=1\linewidth]{\figpath 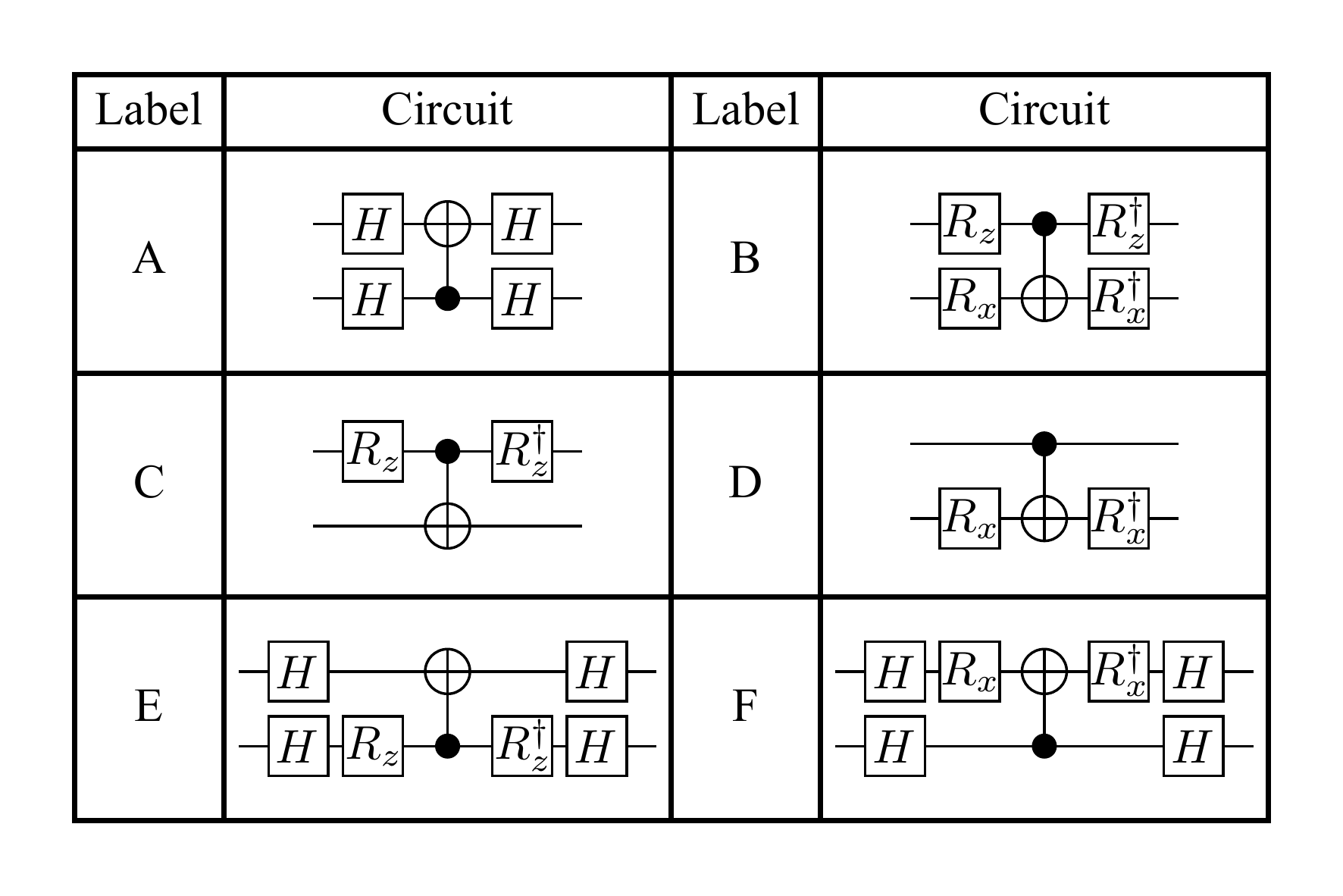}
\caption{
Modified controlled-NOT gates. All these six circuits are equivalent to the original controlled-NOT gate if circuits are error-free. Here, $R_z = \exp^{i\frac{\pi}{4}Z}$ and $R_x = \exp^{i\frac{\pi}{4}X}$. 
}
\label{fig:circuit_CNOT}
\end{figure}

Before explaining our construction, we introduce a universal way to construct a $n$-qubit parity-check circuit (which may not be fault-tolerant), as shown in Fig.~\ref{fig:circuit_CandP}(b). To measure $g = Z\otimes Z\otimes \cdots \otimes Z$ of $n$ qubits, an ancillary qubit is initialised in the state $\ket{0}$, then a sequence of controlled-NOT gates are performed on every data qubit and the ancillary qubit. The ancillary qubit is measured in the computational basis. If the outcome is $\ket{0}$, the outcome of $g$ is the eigenvalue $\mu = +1$; and if the outcome is $\ket{1}$, $\mu = -1$. 

\begin{figure}[tbp]
\centering
\includegraphics[width=1\linewidth]{\figpath 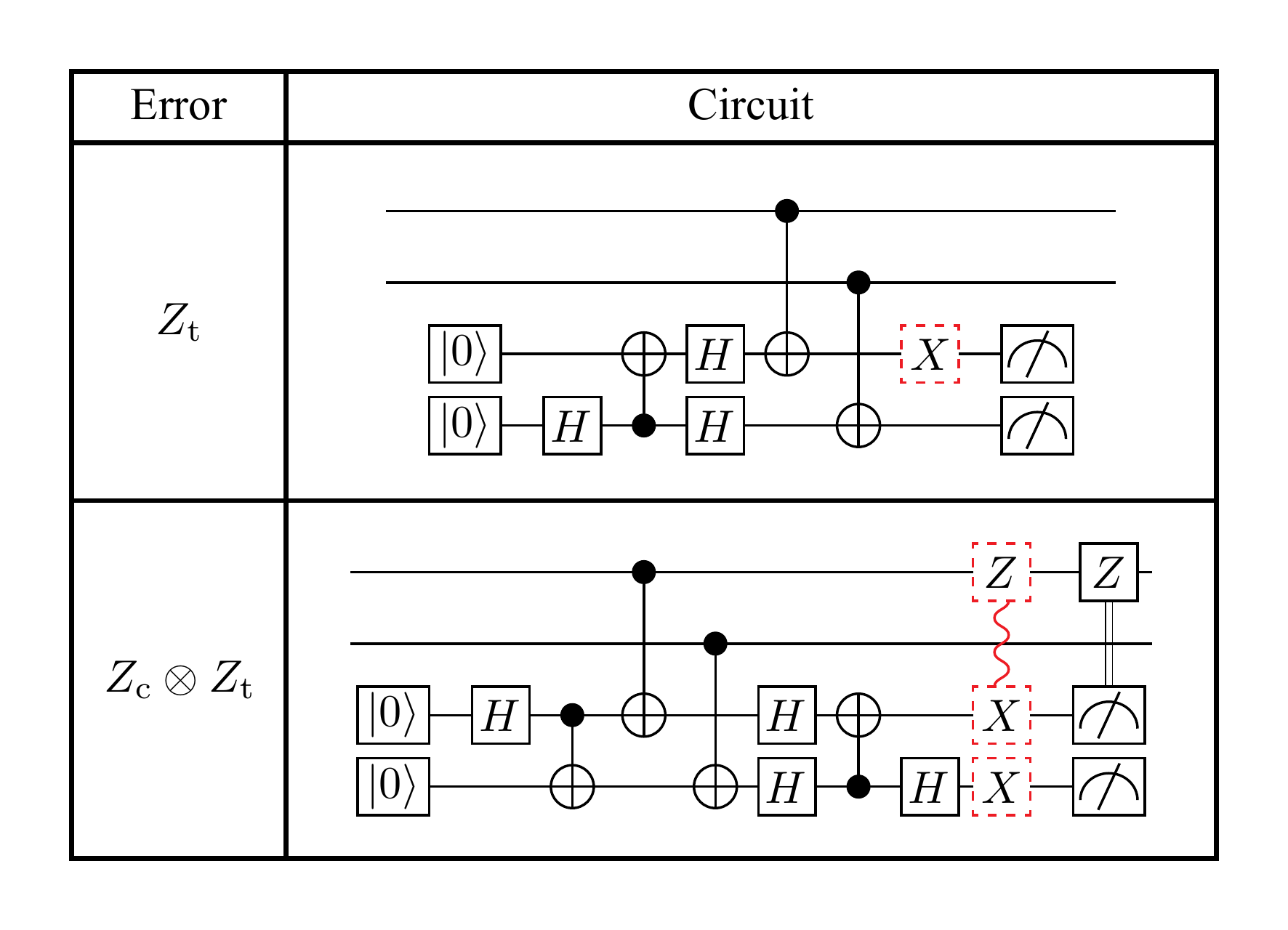}
\caption{
The 1-channel efficient circuits for $Z_{\rm t}$ and $Z_{\rm c}\otimes Z_{\rm t}$. The dashed red square denotes the error that may happen. Two errors connected by the wavy line always occur simultaneously. Circuits for $X_{\rm c}$, $Y_{\rm t}$, and $Y_{\rm c}$ ($X_{\rm c}\otimes X_{\rm t}$, $Z_{\rm c}\otimes Y_{\rm t}$ and $Y_{\rm c}\otimes X_{\rm t}$) can be obtained by replacing the original controlled NOT gate in the circuit for $Z_{\rm t}$ ($Z_{\rm c}\otimes Z_{\rm t}$) with modified controlled-NOT gates A, D, and E in Fig.~\ref{fig:circuit_CNOT}, respectively. 
}
\label{fig:circuit_1cB}
\end{figure}

Parity-check circuits constructed as in Fig.~\ref{fig:circuit_CandP}(b) are robust to $X_{\rm t}$ errors. Because the ancillary qubit is always the target qubit in controlled-NOT gates, $X_{\rm t}$ errors commute with all later gates and accumulate on the ancillary qubit. These errors cause incorrect outcome of the parity check. If the number of $X_{\rm t}$ errors that occur in the circuit is odd, the measurement outcome is flipped: When the state of data qubits is projected into $[(\openone+\mu g)/2]\rho$, the outcome is $-\mu$ rather than the correct value $\mu$. $X_{\rm t}$ errors do not change the state of data qubits. Therefore, under the condition that the probability of such a measurement error is lower than $1/2$, we can correct the error by repeating the measurement and determine the eventual outcome by the majority vote. Suppose that the outcome is $+1$ in $N_+$ measurements and $-1$ in $N_-$ measurements, the eventual outcome is $+1$ if $N_+>N_-$ and $-1$ if $N_->N_+$. 

\begin{figure}[tbp]
\centering
\includegraphics[width=1\linewidth]{\figpath 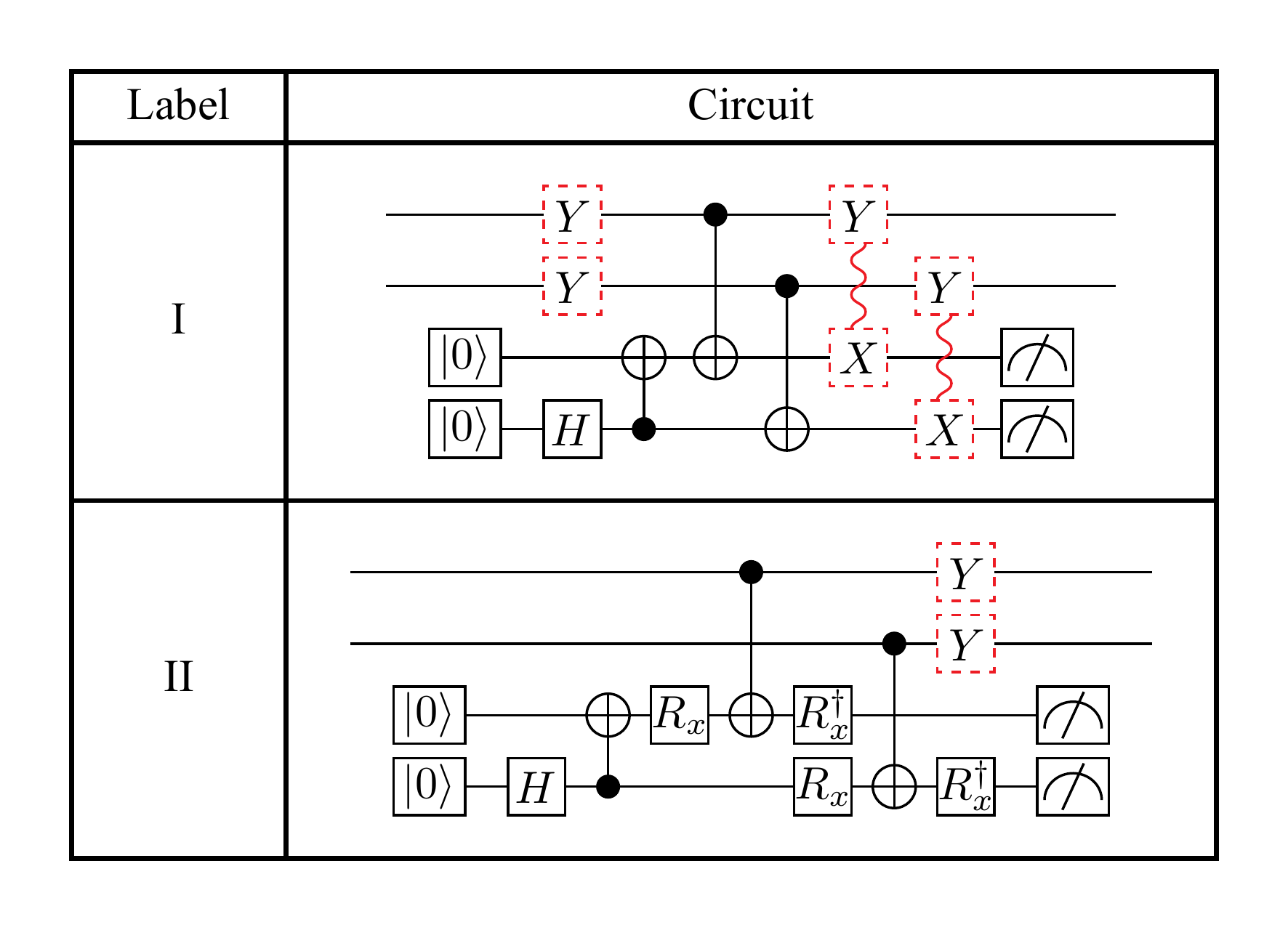}
\caption{
The 1-channel efficient circuits for $Y_{\rm c}\otimes Y_{\rm t}$. The dashed red square denotes the error that may happen. Two errors connected by the wavy line always occur simultaneously. Circuits for $X_{\rm c}\otimes Z_{\rm t}$, $X_{\rm c}\otimes Y_{\rm t}$ and $Y_{\rm c}\otimes Z_{\rm t}$ can be obtained by using modified controlled-NOT gates B, C, and D in Fig.~\ref{fig:circuit_CNOT}, respectively. 
}
\label{fig:circuit_1cC}
\end{figure}

\subsection{Two-qubit parity-check circuits}
\label{sec:2qpcc}

We first consider the 1-channel case, i.e.~only one of 15 Pauli error channels in the controlled-NOT gate is dominant. We construct two-qubit parity-check circuits that are efficient in the error correction for each Pauli error channel as follows. The 15 channels can be divided in to four groups. 

\begin{figure}[tbp]
\centering
\includegraphics[width=1\linewidth]{\figpath 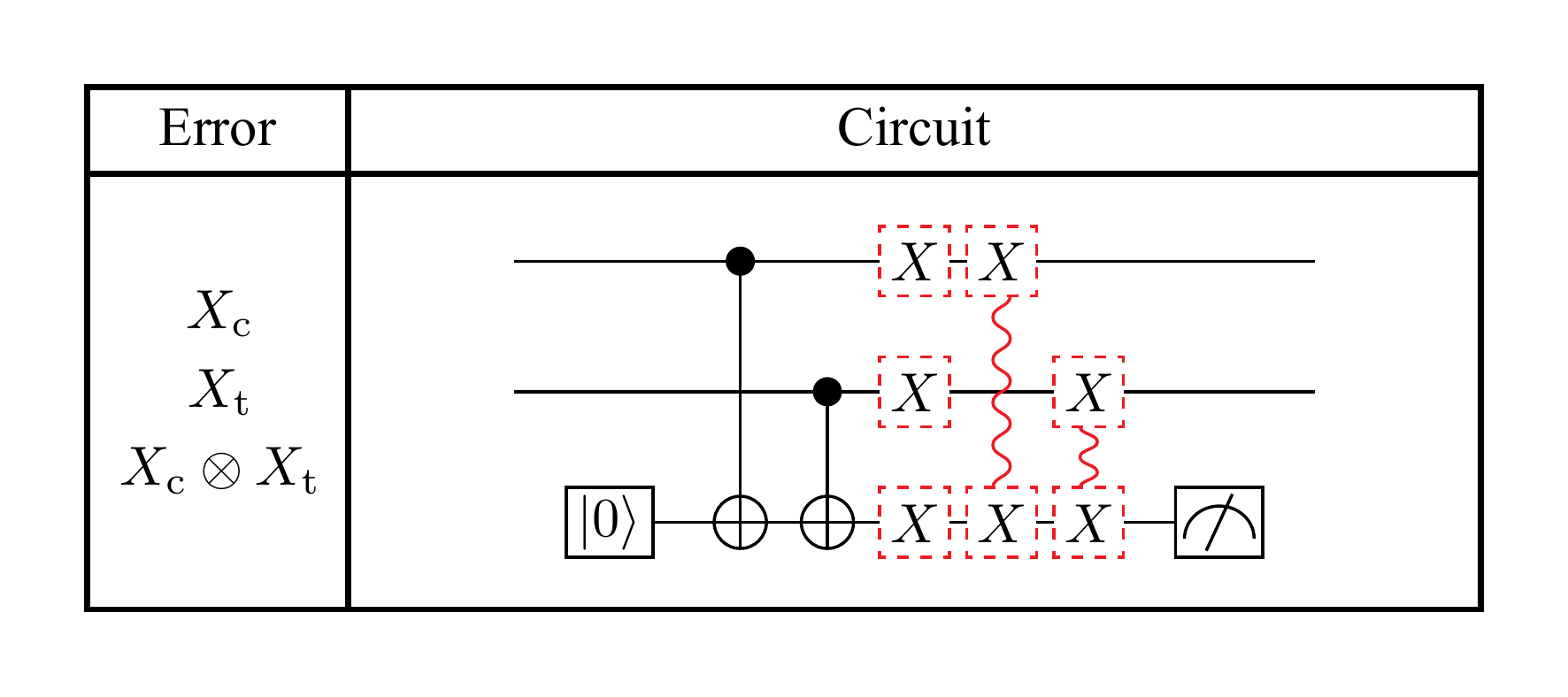}
\caption{
The 3-channel efficient circuit for $\{X_{\rm c}, X_{\rm t}, X_{\rm c}\otimes X_{\rm t}\}$. The dashed red square denotes the error that may happen. Two errors connected by the wavy line always occur simultaneously. Circuits for $\{Z_{\rm c}, Z_{\rm t}, Z_{\rm c}\otimes Z_{\rm t}\}$, $\{Y_{\rm c}, X_{\rm t}, Y_{\rm c}\otimes X_{\rm t}\}$ and $\{Z_{\rm c}, Y_{\rm t}, Z_{\rm c}\otimes Y_{\rm t}\}$ can be obtained by using modified controlled-NOT gates A, C, and F in Fig.~\ref{fig:circuit_CNOT}, respectively. 
}
\label{fig:circuit_3c}
\end{figure}

The first group includes three channels: $X_{\rm t}$, $Z_{\rm c}$ and $Z_{\rm c}\otimes X_{\rm t}$. The corresponding circuits are shown in Fig.~\ref{fig:circuit_1cA}. The circuit for $X_{\rm t}$ is constructed following the approach in Fig.~\ref{fig:circuit_CandP}(b). As we have discussed, $X_{\rm t}$ errors can be efficiently corrected by repeating the parity check. The circuit for $Z_{\rm c}$ is similar. By applying Hadamard gates before and after the controlled-NOT gate, we can realise a controlled-NOT gate with the control qubit and target qubit exchanged (See circuit-B in Fig.~\ref{fig:circuit_CNOT}). With the Hadamard gates, $Z_{\rm c}$ errors are converted into $X_{\rm t}$ errors, therefore they can be efficiently corrected. For $Z_{\rm c}\otimes X_{\rm t}$, the circuit is the same as $X_{\rm t}$. The $Z_{\rm c}\otimes X_{\rm t}$ error causes the measurement error and a phase-flip error on a data qubit. If we repeat the parity check, the phase-flip error does not change measurement outcomes. Therefore, we still can efficiently correct the measurement error. Given the eventual outcome, we can find out how many measurement errors have happened, which is the smaller number in $N_+$ and $N_-$. Because the measurement error and the phase-flip error are associated, we can correct the phase-flip error given the measurement error number. We apply a $Z$ gate on one of two data qubits if and only if the number is odd. We remark that phase-flip errors on two data qubits are equivalent, because the state of data qubits after the parity check is an eigenstate of $Z\otimes Z$. 

\begin{figure}[tbp]
\centering
\includegraphics[width=1\linewidth]{\figpath 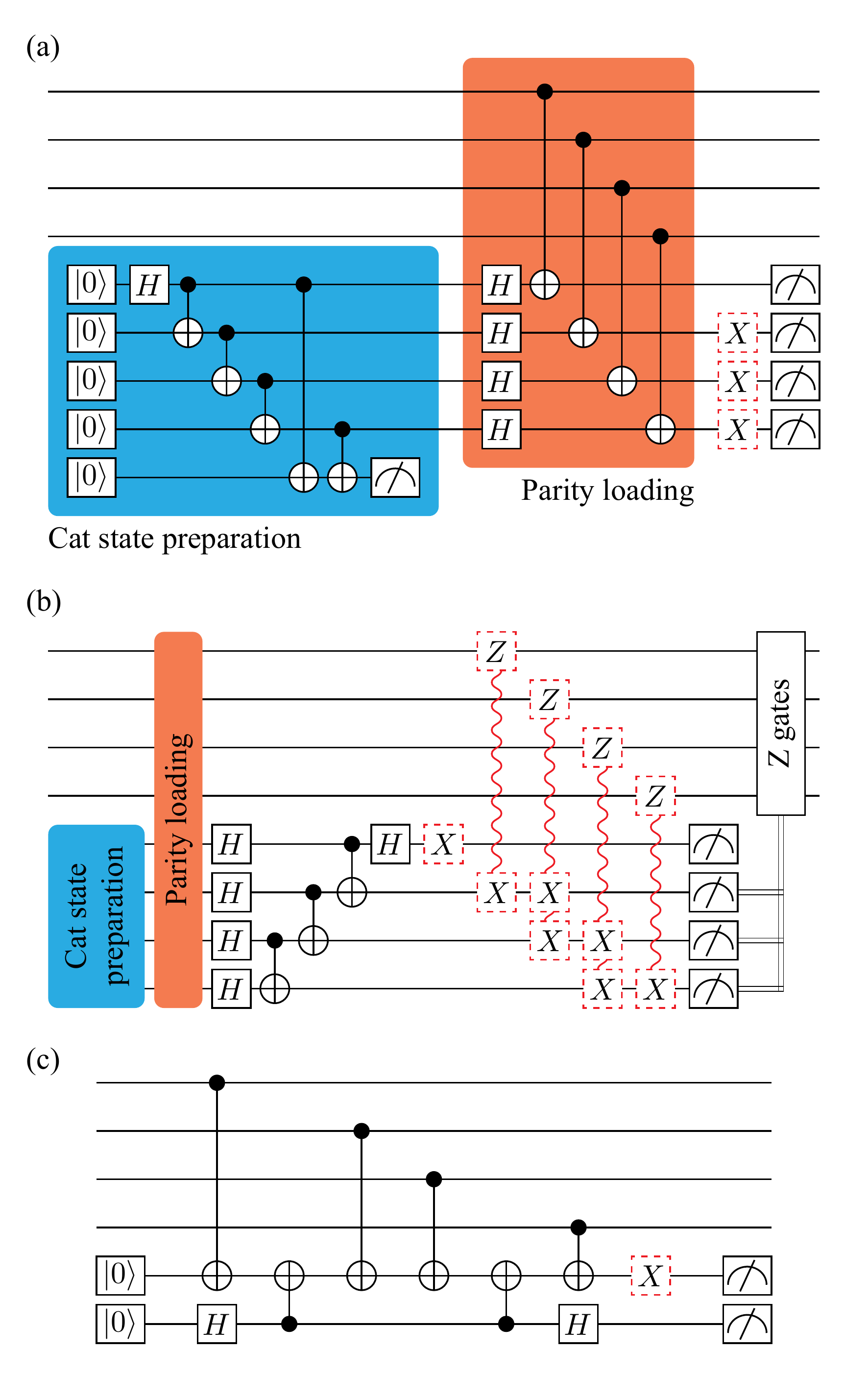}
\caption{
Fault-tolerant parity check circuits of the Steane code. The dashed red square denotes the error that may happen. Two or three errors connected by the wavy line always occur simultaneously. (a) The circuit is efficient for the error channel $Z_{\rm t}$. (b) The circuit is efficient for the error channel $Z_{\rm c}\otimes Z_{\rm t}$. Circuits for $X_{\rm c}$, $Y_{\rm t}$, and $Y_{\rm c}$ ($X_{\rm c}\otimes X_{\rm t}$, $Z_{\rm c}\otimes Y_{\rm t}$ and $Y_{\rm c}\otimes X_{\rm t}$) can be obtained by replacing the original controlled NOT gate in the circuit for $Z_{\rm t}$ ($Z_{\rm c}\otimes Z_{\rm t}$) with modified controlled-NOT gates A, D, and E in Fig.~\ref{fig:circuit_CNOT}, respectively. 
}
\label{fig:circuit_Steane}
\end{figure}

The second group includes four channels: $Z_{\rm t}$, $X_{\rm c}$, $Y_{\rm t}$, and $Y_{\rm c}$. The circuit for $Z_{\rm t}$ has two ancillary qubits, which are prepared in the Bell state $\frac{1}{\sqrt{2}}(\ket{00}+\ket{11})$ using a Hadamard gate and a controlled-NOT gate, as shown in Fig.~\ref{fig:circuit_1cB}(a). Two ancillary qubits are measured at the end of the circuit. If the outcome of ancillary qubits is $\ket{00}$ or $\ket{11}$, the outcome of the parity check is $+1$; and if the outcome of ancillary qubits is $\ket{01}$ or $\ket{10}$, the outcome of the parity check is $-1$. We can find that the $Z_{\rm t}$ error of the first controlled-NOT gate leads to a measurement error, and $Z_{\rm t}$ errors of other two controlled-NOT gates do not have any effect on neither data qubits nor the outcome. For $X_{\rm c}$, $Y_{\rm t}$, and $Y_{\rm c}$ errors, they can be converted into $Z_{\rm t}$ errors by applying appropriate single-qubit gates before and after the controlled-NOT gate, according to circuits A, D, and E in Fig.~\ref{fig:circuit_CNOT}, respectively. 

The third group also includes four channels: $Z_{\rm c}\otimes Z_{\rm t}$, $X_{\rm c}\otimes X_{\rm t}$, $Z_{\rm c}\otimes Y_{\rm t}$ and $Y_{\rm c}\otimes X_{\rm t}$. Similar to $Z_{\rm t}$, the circuit for $Z_{\rm c}\otimes Z_{\rm t}$ also uses two ancillary qubits prepared in the Bell state, however the readout strategy is different, as shown in Fig.~\ref{fig:circuit_1cB}(b). When the circuit is error-free, the outcome of the upper ancillary qubit is always $\ket{0}$, and the outcome of the lower ancillary qubit indicates the outcome of the parity check. If the outcome of the lower ancillary qubit is $\ket{0}$, the outcome of the parity check is $+1$; and if the outcome of the lower ancillary qubit is $\ket{1}$, the outcome of the parity check is $-1$. We can find that the $Z_{\rm c}\otimes Z_{\rm t}$ error of the first controlled-NOT gate causes a pair of phase-flip errors on data qubits. Because the state of data qubits after the parity check is an eigenstate of $Z\otimes Z$, such a pair of phase-flip errors is trivial. The $Z_{\rm c}\otimes Z_{\rm t}$ error in the second controlled-NOT gate causes a phase-flip error on a data qubit and a measurement error on the upper ancillary qubit. It is similar for the third controlled-NOT gate. Therefore, the upper ancillary qubit can be used to detect errors. If the outcome of the upper ancillary qubit is $\ket{1}$ rather than $\ket{0}$, there must be a phase-flip error on one of two data qubits, then we can correct it by applying a $Z$ gate. We remark that phase-flip errors on two data qubits are equivalent. The fourth controlled-NOT gate leads to the measurement error on the lower ancillary qubit, i.e.~measurement error of the parity check. For $X_{\rm c}\otimes X_{\rm t}$, $Z_{\rm c}\otimes Y_{\rm t}$ and $Y_{\rm c}\otimes X_{\rm t}$ errors, they can be converted into $Z_{\rm c}\otimes Z_{\rm t}$ errors by applying appropriate single-qubit gates before and after the controlled-NOT gate, according to circuits A, D, and E in Fig.~\ref{fig:circuit_CNOT}, respectively. 

The fourth group includes the last four channels: $Y_{\rm c}\otimes Y_{\rm t}$, $X_{\rm c}\otimes Z_{\rm t}$, $X_{\rm c}\otimes Y_{\rm t}$ and $Y_{\rm c}\otimes Z_{\rm t}$. For the other three groups, dominant errors in controlled-NOT gates result in measurement errors of the parity check, by using corresponding circuits. The fourth group is different, and we cannot find such measurement-error circuits. For channels in the fourth group, we find circuits that dominant errors in controlled-NOT gates result in bit-flip errors on data qubits, which can be efficiently corrected using the repetition code (see Sec.~\ref{sec:1c-RC}). We introduce two circuits for $Z_{\rm c}\otimes Z_{\rm t}$, which are similar to the circuit for $Z_{\rm t}$, as shown in Fig.~\ref{fig:circuit_1cC}. In the circuit I, we can find that the $Z_{\rm c}\otimes Z_{\rm t}$ error of the first controlled-NOT gate is trivial, and errors of the other two controlled-NOT gates cause the measurement error and a Y error on one of two data qubits. Such correlated errors are equivalent to Y errors occurring before the parity check. In the circuit II, we can find that the $Z_{\rm c}\otimes Z_{\rm t}$ error of the first controlled-NOT gate is still trivial, but errors of the other two controlled-NOT gates only cause Y errors on data qubits, i.e.~Y errors occurring after the parity check. For $X_{\rm c}\otimes Z_{\rm t}$, $X_{\rm c}\otimes Y_{\rm t}$ and $Y_{\rm c}\otimes Z_{\rm t}$ errors, they can be converted into $Y_{\rm c}\otimes Y_{\rm t}$ errors by applying appropriate single-qubit gates before and after the controlled-NOT gate, according to circuits B, C, and D in Fig.~\ref{fig:circuit_CNOT}, respectively. 

Now, we consider the 3-channel case, i.e.~three of 15 Pauli error channels in the controlled-NOT gate are dominant. The three channels are $\{X_{\rm c}, X_{\rm t}, X_{\rm c}\otimes X_{\rm t}\}$. The circuit for this set of three channels is the same as the circuit for $X_{\rm t}$, as shown in Fig.~\ref{fig:circuit_3c}. The $X_{\rm c}$ errors in controlled-NOT gates cause bit-flip errors on data qubits, the $X_{\rm t}$ errors cause measurement errors, and $X_{\rm c}\otimes X_{\rm t}$ errors cause correlated errors. These errors can be efficiently corrected using the repetition code (see Sec.~\ref{sec:3c}). Similar three-error sets are $\{Z_{\rm c}, Z_{\rm t}, Z_{\rm c}\otimes Z_{\rm t}\}$, $\{Y_{\rm c}, X_{\rm t}, Y_{\rm c}\otimes X_{\rm t}\}$ and $\{Z_{\rm c}, Y_{\rm t}, Z_{\rm c}\otimes Y_{\rm t}\}$, they can be converted into $\{X_{\rm c}, X_{\rm t}, X_{\rm c}\otimes X_{\rm t}\}$ by applying appropriate single-qubit gates before and after the controlled-NOT gate, according to circuits A, C, and F in Fig.~\ref{fig:circuit_CNOT}, respectively. 

\subsection{Four-qubit parity-check circuits}

Four-qubit parity checks are used in many quantum error correction codes, e.g.~the surface code and Steane code. For the surface code, we can construct the parity check circuit according to Fig.~\ref{fig:circuit_CandP}(b), then the error correction is efficient if the $X_{\rm t}$ (or $Z_{\rm c}$, up to Hadamard gates) channel is dominant in controlled-NOT-gate errors. However, for the Steane code, because the code is compact with a distance of $3$, the circuit constructed according to Fig.~\ref{fig:circuit_CandP}(b) is not fault-tolerant. In Fig.~\ref{fig:circuit_Steane}, we show three fault-tolerant circuit of the Steane code, reported in Refs.~\cite{ Xu2018, Chao2018,DiVincenzo2007}, respectively. Each of the circuits is efficient for a category of 1-channel errors, as follows. 

The circuit in Fig.~\ref{fig:circuit_Steane}(a) is formed by two parts. The first part prepares a cat state. Errors in the cat state are detected by the ancillary qubit on the bottom. If any error is detected at this stage, the cat state is discarded, and the state preparation restarts, which is repeated until the cat state is successfully prepared. The second part loads the parity of four data qubits onto the cat state. Finally, four ancillary qubits are measured in the computational basis, and the value of the parity is the eigenvalue of $Z\otimes Z\otimes Z\otimes Z$ of the four ancillary qubits. We can find that $Z_{\rm t}$ errors of the first four controlled-NOT gates lead to measurement errors, and $Z_{\rm t}$ errors of other controlled-NOT gates are trivial. Therefore, this circuit is efficient when $Z_{\rm t}$ errors are dominant. It is similar for $X_{\rm c}$, $Y_{\rm t}$ and $Y_{\rm c}$ errors, which can be converted into $Z_{\rm t}$ errors by applying appropriate single-qubit gates before and after the controlled-NOT gate, according to circuits A, D, and E in Fig.~\ref{fig:circuit_CNOT}, respectively. 

The second circuit is similar to the first circuit but uses a different readout strategy, see Fig.~\ref{fig:circuit_Steane}(b). Here, only the outcome of the top ancillary qubit indicates the parity of four data qubits. If the outcome of the top ancillary qubit is $\ket{0}$ or $\ket{1}$, the parity is $+1$ or $-1$, respectively. Outcomes of other three qubits are used for detecting errors. When the circuit is error-free, outcomes of these three qubits are always $\ket{0}$. For $Z_{\rm c}\otimes Z_{\rm t}$ errors in the cat state preparation and the last three controlled-NOT gates for readout, we can find that they only lead to measurement errors on the top ancillary qubit. However, $Z_{\rm c}\otimes Z_{\rm t}$ errors in the parity loading can cause correlated errors on data qubits and the other three ancillary qubits. By measuring the lower three ancillary qubits, these phase-flip errors on data qubits can be corrected, and only measurement errors of the parity are left. Therefore, this circuit is efficient when $Z_{\rm c}\otimes Z_{\rm t}$ errors are dominant. It is similar for $X_{\rm c}\otimes X_{\rm t}$, $Z_{\rm c}\otimes Y_{\rm t}$ and $Y_{\rm c}\otimes X_{\rm t}$ errors, which can be converted into $Z_{\rm c}\otimes Z_{\rm t}$ errors by applying appropriate single-qubit gates before and after the controlled-NOT gate, according to circuits A, D, and E in Fig.~\ref{fig:circuit_CNOT}, respectively. 

The third circuit is constructed using a different approach, which uses only two ancillary qubits, as shown in Fig.~\ref{fig:circuit_Steane}(c). The outcome of the upper ancillary qubit indicates the parity of four data qubits, and the lower qubit is used to detect weight-2 errors. Both of the two outcomes will be used in the later error correction. We can find that $X_{\rm t}$ errors only lead to measurement errors on the upper ancillary qubit, i.e.~incorrect outcome of the parity check. Therefore, this circuit is efficient when $X_{\rm t}$ errors are dominant. It is similar for $Z_{\rm c}$ errors, which can be converted into $X_{\rm t}$ errors by applying Hadamard gates before and after the controlled-NOT gate, according to the circuit A in Fig.~\ref{fig:circuit_CNOT}. 

For other five 1-channel errors, we have not found efficient fault-tolerant circuits. 

\section{Logical error rate}
\label{sec:error_rate}

\begin{figure*}[tbp]
\centering
\includegraphics[width=1\linewidth]{\figpath 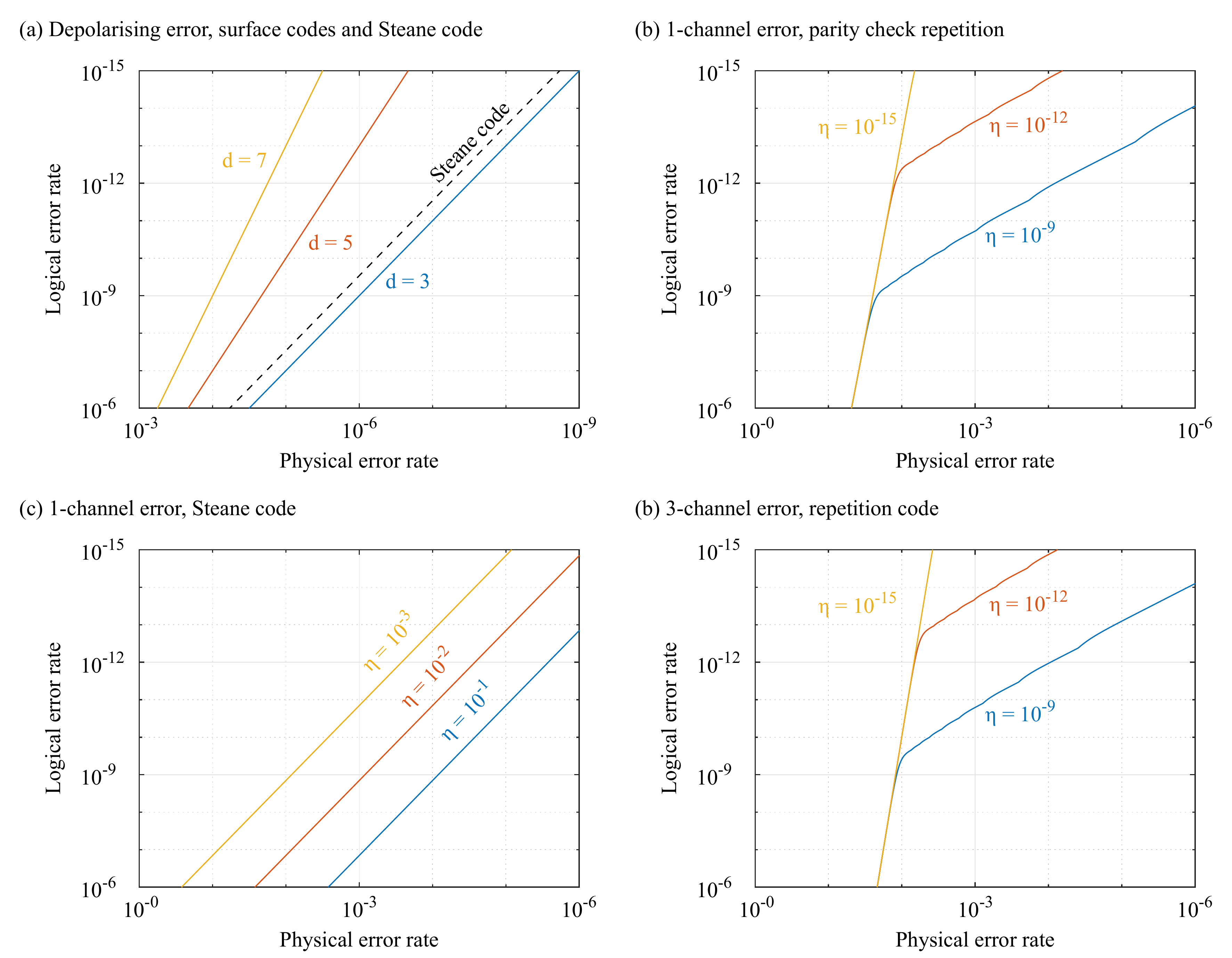}
\caption{
The logical error rate versus the physical error. In (b) and (d), the logical error rate is obtained by optimising the code distance $d$. For the parity check repetition, $d = 3,5,\ldots,21$; for the repetition code, $d = 4,6,\ldots,22$. $\eta = \epsilon/p$ is the bias ratio. 
}
\label{fig:error_rate}
\end{figure*}

The logical error rate is the probability of errors occurring on logical qubits. In the case of 1-channel errors in the first three groups, we can use circuits in Figs.~\ref{fig:circuit_1cA}~and~\ref{fig:circuit_1cB} to implement the two-qubit parity check. Using these circuits, the dominant errors are measurement errors of the parity check, which can be corrected by repeating the parity check without using any error correction code. With the error-corrected parity check, we can realise universal quantum computation. Therefore, in this scenario, each physical qubit is a logical qubit, and we take the error rate of the error-corrected parity check as the logical error rate. 

In the case that an error correction code is used,we repeatedly measure stabiliser generators using parity checks, which are the fundamental operations of fault-tolerant quantum computation. These stabiliser measurements can correct errors generated by themselves and preserve the logical information. To actively operate logical qubits, i.e.~implement logical gates, we need to modify the periodic stabiliser measurement circuit, e.g.~inserting transversal gates between two stabiliser cycles. These modifications may introduce additional errors, which will be corrected by subsequent stabiliser measurements. A logical gate may contain several rounds of stabiliser measurement, for example the surface code. Because of the fundamental role of stabiliser measurements, we take the rate of logical errors per round of stabiliser measurements as the measure of performance. 

\section{Error correction for depolarising errors}
\label{sec:depol}

When the 15 Pauli error channels have similar rates, such as in the depolarizing error model, we need to use quantum error correction codes, e.g.~the surface code and Steane code, to correct the errors. For the surface code, the logical error rate per round of stabiliser measurements is~\cite{Fowler2013} 
\begin{eqnarray}
p_{\rm L} \simeq 0.1(100p)^{\frac{d+1}{2}},
\end{eqnarray}
where $p = 1 - P_{I_{\rm c},I_{\rm t}}$ is the total error rate of each controlled-NOT gate, and $d$ is the code distance. As a comparison to biased error channels, we plot the logical error rate versus the physical error rate for the surface code in Fig.~\ref{fig:error_rate}(a). 

The Steane code error correction using the circuit in Fig.~\ref{fig:circuit_Steane}(a) are simulated numerically, and the logical error rate is plotted in Fig.~\ref{fig:fitting_Steane}. The decoder will be discussed in Sec.~\ref{sec:Steane}. We fit the logical error rate using the formula 
\begin{eqnarray}
p_{\rm L} = (\alpha p)^2, 
\end{eqnarray}
where fitting parameter is found to be $\alpha = 17.0914$ with the standard deviation $\sigma_\alpha = 0.8915$. The logical error rate versus the physical error rate for the Steane code, according to the fitting formula, is also plotted in Fig.~\ref{fig:error_rate}(a). We can find that the performance of the Steane code is slightly better than the surface code with the distance $d = 3$. However, it is not a fair comparison because only two-qubit-gate errors are taken into account in our numerical simulations. 

\section{Error correction for biased 1-channel errors}
\label{sec:1c}

In the 1-channel case, if the error rate of minor errors is on the fault-tolerant level even without using the error correction, we only need to correct dominant errors. In this case, we can use the two-qubit parity check as the building block. For the first three groups of error channels (see Sec.~\ref{sec:2qpcc}), the error can be corrected by repeating the parity check itself, and the error correction encoding is not needed. For the fourth group, dominant errors result in bit-flip errors, but the parity projection has a reliable measurement outcome. Then, we can use the parity projection to implement the classical repetition code to correct the bit-flip errors. If minor errors are not negligible, we can use a quantum error correction code, the Steane code, to correct minor errors, and we use the parity check circuit that is efficient for correcting the dominant error in the error correction. In this section, we discuss all these situations. 

\subsection{Repetition code}
\label{sec:1c-RC}

\begin{figure}[tbp]
\centering
\includegraphics[width=1\linewidth]{\figpath 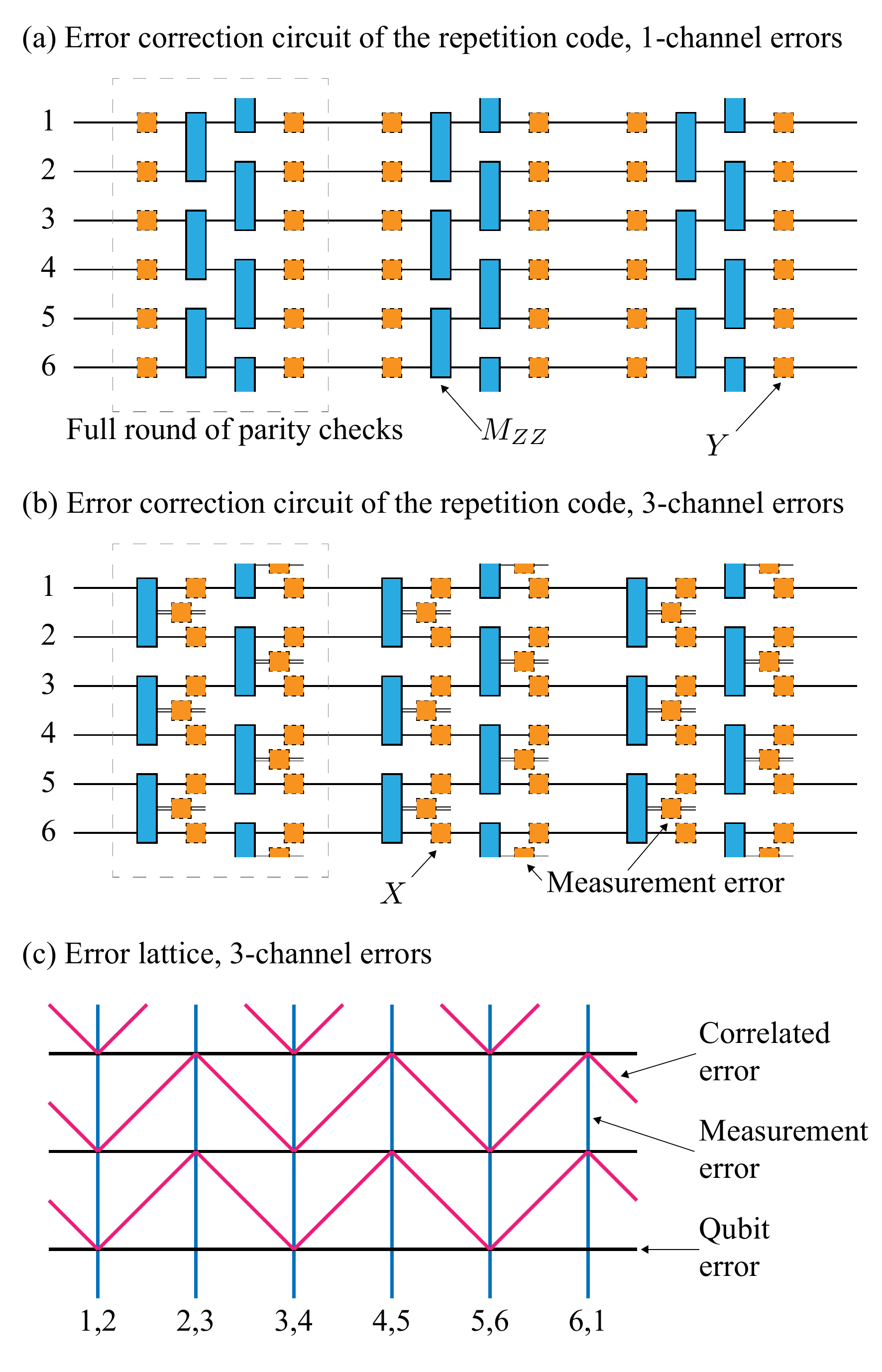}
\caption{
(a) The repetition code error correction in the 1-channel case. The dominant error only causes $Y$ errors on data qubits. 
(b) The repetition code error correction in the 3-channel case. The dominant errors causes $X$ errors on data qubits and measurement errors. 
(c) The error correction lattice for the 3-channel repetition code error correction. 
}
\label{fig:code_R.pdf}
\end{figure}

For the first three groups of error channels (see Sec.~\ref{sec:2qpcc}), the dominant error only results in the incorrect outcome of the parity check, which can be corrected by repeating the parity check. If the parity check is repeated $d$ times, the eventual measurement outcome is correct if the number of measurement errors is not larger than $\lfloor (d-1)/2 \rfloor$. Therefore, the logical error rate of the parity check is 
\begin{eqnarray}
p_{\rm L} \simeq \sum_{n = \lfloor (d-1)/2 \rfloor+1}^d \binom{d}{n} (Ap)^n (1-Ap)^{d-n} + dB\frac{\epsilon}{14},~~
\label{eq:RC1}
\end{eqnarray}
where $p$ is the error rate of the dominant error in one controlled-NOT gate, and $\epsilon$ is the error rate of minor errors, i.e.~the error rate of each minor error is $\epsilon/14$. Here, $A$ is the number of dominant error channels that can cause the measurement error, and $B$ is the number of minor error channels that cause errors on two data qubits. For example, for the $X_{\rm t}$ error (see Fig.~\ref{fig:circuit_1cA}), the $X_{\rm t}$ errors in both of two controlled-NOT gates result in the measurement error, therefore $A = 2$. In the total $28$ minor error channels, $24$ of them can cause errors on data qubits, two of them cause the measurement error (which are neglected compared with dominant error channels), and two of them are trivial, therefore $B = 24$. To obtain Eq.~(\ref{eq:RC1}), we have assumed that $p \ll 1$ and $\epsilon \ll p$. The logical error rate for the $X_{\rm t}$-dominant error model is plotted in Fig.~\ref{fig:error_rate}(b). It is similar for the other ten error channels in the first three groups. 

For the fourth group, the dominant error results in $Y$ errors on data qubits but does not affect the outcome of the parity projection. Using the repetition code, we can correct these $Y$ errors. In the repetition code of distance $d$, two logical states are $\ket{0_{\rm L}} = \ket{0}^{\otimes d}$ and $\ket{1_{\rm L}} = \ket{1}^{\otimes d}$, and stabiliser generators are $I^{\otimes i-1}\otimes Z_i\otimes Z_{i+1} \otimes I^{d-i-1}$, where $Z_i$ is the Pauli operator on the $i$-th qubit. The circuit for measuring stabiliser generators are shown in Fig.~\ref{fig:code_R.pdf}. Each full round of parity checks is formed by two layers of parity checks. For the first layer, we use the circuit-I in Fig.~\ref{fig:circuit_1cC} such that $Y$ errors are effectively placed before the parity checks; for the second layer, we use the circuit-II in Fig.~\ref{fig:circuit_1cC} such that $Y$ errors are placed after the parity checks. In this way, all $Y$ errors only appear between two full rounds of parity checks. Then, we can compare outcomes of the two full rounds and correct $Y$ errors. The $Y$ errors can be successfully corrected if their number is not larger than $\lfloor (d-1)/2 \rfloor$. We note that other errors on data qubits and measurement errors cannot be corrected. The logical error rate can be expressed as in  Eq.~(\ref{eq:RC1}), where $A$ is the number of dominant error channels that cause the $Y$ error on a data qubit, and $B$ is the number of minor error channels that cause other errors. 

\subsection{Steane code}
\label{sec:Steane}

\begin{figure}[tbp]
\centering
\includegraphics[width=1\linewidth]{\figpath 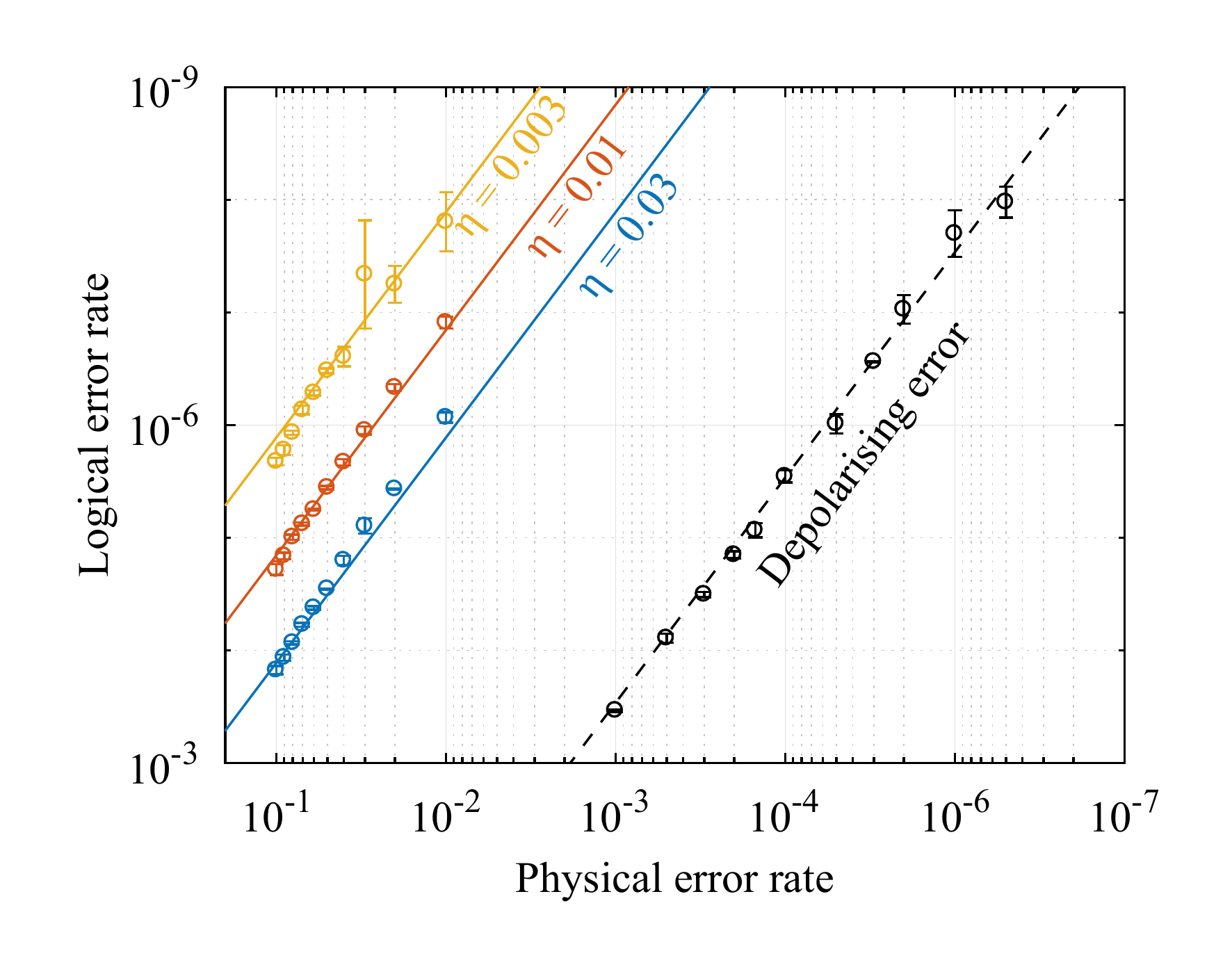}
\caption{
Fitting (lines) to numerical-simulation data (scatters) for the Steane code. 
}
\label{fig:fitting_Steane}
\end{figure}

For the Steane code, we take the parity-check circuit in Fig.~\ref{fig:circuit_Steane}(a) as an example, which can efficiently correct $Z_{\rm t}$ errors in the one-channel case. It is similar for other circuits in Fig.~\ref{fig:circuit_Steane}. In terms of the error correction decoder, we use a message-passing scheme that maximises the chance of successfully correcting errors~\cite{Poulin2006, Stephens2009}. Before each round of parity checks, there is an input list of all possible error configurations and their probabilities. After the parity checks, the list is updated, because new errors are introduced by the parity-check operations. With the measurement outcomes of party checks, the probabilities in the list are updated again: only error configurations resulting in the outcome pattern survive, their probabilities are renormalised, and probabilities of all other error configurations are set to zero. In this way, we obtain the posterior distribution. With the posterior distribution, the correction operation is performed according to the most likely error configuration. Then, we need to update the list once more to take into account the effect of correction operations. The output list is used as the input list for the next round. 

We compute the logical error rates of the Steane code using Monte Carlo simulations, and plot the results in Fig.~\ref{fig:fitting_Steane}. In the simulations, we take $\epsilon/p = 0.003,0.01,0.03$. Then, we fit the logical error rates using the formula 
\begin{eqnarray}
p_{\rm L} = (\beta \epsilon)^2, 
\end{eqnarray}
where fitting parameter is found to be $\beta = 3.7815$ with the standard deviation $\sigma_\beta = 0.0985$. Using the fitting formula, we compute the logical error rates for $\epsilon/p = 0.001,0.01,0.1$, which are plotted in Fig.~\ref{fig:error_rate}(c). We can find that, using the Steane code, a much higher $\epsilon$ is tolerable compared with the repetition code. This is because the Steane code can correct general errors up to the code distance of $3$. 

\section{Error correction for biased 3-channel errors}
\label{sec:3c}

\begin{figure}[tbp]
\centering
\includegraphics[width=1\linewidth]{\figpath 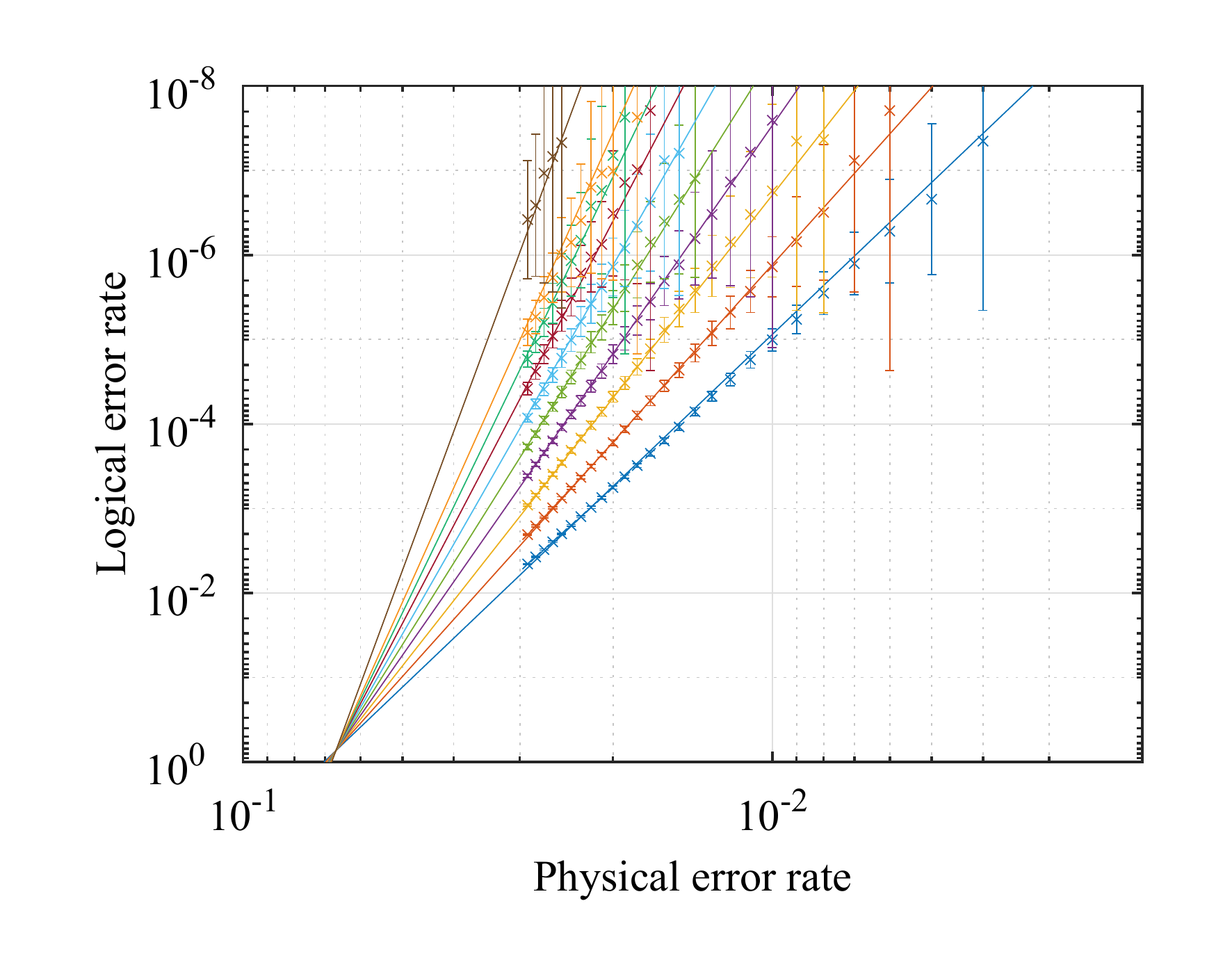}
\caption{
Fitting (lines) to numerical-simulation data (scatters) for the repetition code. For each curve from bottom to top, the code distance is $d = 10,12,14,16,18,20,22,24,26,32$, respe
}
\label{fig:fitting_Repetition}
\end{figure}

In the 3-channel case, see the circuit in Fig.~\ref{fig:circuit_3c}, the two-qubit parity check may cause both qubit errors and measurement error. To correct these errors, we need to use the repetition code, similar to the fourth group of error channels in the 1-channel case. As shown in Fig.~\ref{fig:code_R.pdf}(b), we use the circuit in Fig.~\ref{fig:circuit_3c} in both two layers of parity checks in a full round. Measurement errors are corrected in a way similar to the surface code~\cite{Fowler2012, Dennis2002}. We use a two-dimensional error-correction lattice to represent errors, on which the vertical edges represent measurement errors. Errors are identified using the minimum-weight perfect matching algorithm~\cite{Kolmogorov2009}. The logical error rates are computed using numerical simulations assuming the three dominant errors have the same error rate $p/3$, and the rate of other errors is zero. The results are shown in Fig.~\ref{fig:fitting_Repetition}. We fit the logical error rates using the formula 
\begin{eqnarray}
p_{\rm L} = p_{0}(p/p_{\rm th})^{d/2+\delta}, 
\end{eqnarray}
where fitting parameters are found to be $p_0 = 0.7335$, $p_{\rm th} = 0.0668$ and $\delta = 0.9743$ with the standard deviations $\sigma_{p_0} = 0.0856$, $\sigma_{p_{\rm th}} = 0.0005$ and $\sigma_{p_0} = 0.0521$, respectively. Here, $p_{\rm th}$ is the threshold of the code. 

Similar to the repetition code in the 1-channel case, when the error rate of minor errors is nonzero, the logical error rate is 
\begin{eqnarray}
p_{\rm L} \simeq p_{0}(p/p_{\rm th})^{d/2+\delta} + dB\frac{\epsilon}{12},~~
\label{eq:RC3}
\end{eqnarray}
where $\epsilon$ is the error rate of minor errors, i.e.~the error rate of each minor error is $\epsilon/12$. In the two controlled-NOT gates in Fig.~\ref{fig:circuit_3c}, there are $6$ dominant error channels and $24$ minor error channels. In minor error channels, $16$ of them can cause phase-flip errors on data qubits, $6$ of them only cause bit-flip and measurement errors (which are neglected compared with dominant error channels), and $2$ of them are trivial. Therefore, $B = 16$. The logical error rate for the 3-channel case is plotted in Fig.~\ref{fig:error_rate}(d). 

\section{Summary}
\label{sec:summary}

Our results are summarised in Fig.~\ref{fig:summary}. When the error model is extremely biased, i.e.~the bias ratio $\eta = \epsilon/p = 10^{-12}$, the error correction using the repetition code with a code distance smaller than $22$ is efficient even when the physical error rate is as high as $10^{-3}$. In this case, the difference between 1-channel and 3-channel cases is not significant. Using the Steane code, to achieve the logical error rate of $10^{-15}$, we need the physical error rate about $10^{-6}$ in the 1-channel case with the bias ratio $10^{-2}$, or physical error rate about $10^{-9}$ for depolarising errors. If a larger code, such as the surface code with the distance $d = 7$, is allowed, the fault-tolerant quantum computation can be realised at the physical error rate $10^{-6}$ even for depolarising errors. We can find that the biased error model can significantly reduce the demanding requirement for achieving fault-tolerance, which is similar to the case of the Majorana fermion quantum computation~\cite{Li2016}. To realise the fault-tolerant quantum computation in the deep sub-threshold regime, we may need to either at least reduce the physical error rate to the level of $10^{-6}$ or developing physical systems with a highly biased error model. 

\begin{acknowledgments}
This work is supported by National Natural Science Foundation of China (Grant No. 11875050) and NSAF (Grant No. U1930403). We thank Xiaosi Xu for helpful discussions. 
\end{acknowledgments}


\begin{thebibliography}{9}

\bibitem{Nielsen2010} M. A. Nielsen and I. L. Chuang,
\textit{Quantum Computation and Quantum Information},
Cambridge University Press, Cambridge, (2010).



\bibitem{Fowler2012} A. G. Fowler, M. Mariantoni, J. M. Martinis, and A. N. Cleland,
\textit{Surface codes: Towards practical large-scale quantum computation},
Phys. Rev. A \textbf{86}, 032324 (2012).

\bibitem{OGorman2017} J. O'Gorman and E. T. Campbell,
\textit{Quantum computation with realistic magic state factories},
Phys. Rev. A \textbf{95}, 032338 (2017).



\bibitem{Fowler2009} A. G. Fowler, A. M. Stephens, and P. Groszkowski,
\textit{High-threshold universal quantum computation on the surface code},
Phys. Rev. A \textbf{80}, 052312 (2009).

\bibitem{Wang2011} D. S. Wang, A. G. Fowler, and L. C. L. Hollenberg,
\textit{Surface code quantum computing with error rates over 1\%},
Phys. Rev. A \textbf{83}, 020302(R) (2011).



\bibitem{Barends2014} R. Barends {\it et al.},
\textit{Superconducting quantum circuits at the surface code threshold for fault tolerance},
Nature \textbf{508}, 500 (2014).

\bibitem{Ballance2016} C. J. Ballance, T. P. Harty, N. M. Linke, M. A. Sepiol, and D. M. Lucas,
\textit{High-fidelity quantum logic gates using trapped-ion hyperfine qubits},
Phys. Rev. Lett. \textbf{117}, 060504 (2016).

\bibitem{Gaebler2016} J. P. Gaebler, T. R. Tan, Y. Lin, Y. Wan, R. Bowler, A. C. Keith, S. Glancy, K. Coakley, E. Knill, D. Leibfried, and D. J. Wineland,
\textit{High-fidelity universal gate set for $^9$Be$^+$ ion qubits},
Phys. Rev. Lett. \textbf{117}, 060505 (2016).



\bibitem{Fowler2013} A. G. Fowler, S. J. Devitt and C. Jones,
\textit{Surface code implementation of block code state distillation},
Sci. Rep. \textbf{3}, 1939 (2013).



\bibitem{Google} F. Arute {\it et al.},
\textit{Quantum supremacy using a programmable superconducting processor},
Nature \textbf{574}, 505 (2019).



\bibitem{Li2012} Y. Li and S. C. Benjamin,
\textit{High threshold distributed quantum computing with three-qubit nodes},
New J. Phys. \textbf{14}, 093008 (2012).

\bibitem{Nickerson2013} N. H. Nickerson, Y. Li, and S. C. Benjamin,
\textit{Topological quantum computing with a very noisy network and error rates approaching one percent},
Nat. Commun. \textbf{4}, 1756 (2013).

\bibitem{Monroe2014} C. Monroe, R. Raussendorf, A. Ruthven, K. R. Brown, P. Maunz, L.-M. Duan, and J. Kim,
\textit{Large-scale modular quantum-computer architecture with atomic memory and photonic interconnects},
Phys. Rev. A \textbf{89}, 022317 (2014).



\bibitem{link} https://nqit.ox.ac.uk/content/ion-traps



\bibitem{Nayak2008} C. Nayak, S. H. Simon, A. Stern, M. Freedman, and S. Das Sarma,
\textit{Non-Abelian anyons and topological quantum computation},
Rev. Mod. Phys. \textbf{80}, 1083 (2008).



\bibitem{Aliferis2008} P. Aliferis and J. Preskill,
\textit{Fault-tolerant quantum computation against biased noise}
Phys. Rev. A \textbf{78}, 052331 (2008).

\bibitem{Brooks2013} P. Brooks and J. Preskill,
\textit{Fault-tolerant quantum computation with asymmetric Bacon-Shor codes}
 Phys. Rev. A \textbf{87}, 032310 (2013).

\bibitem{Tuckett2018} D. K. Tuckett, S. D. Bartlett, and S. T. Flammia,
\textit{Ultrahigh error threshold for surface codes with biased noise},
Phys. Rev. Lett. \textbf{120}, 050505 (2018).

\bibitem{Xu2019} X. Xu, Q. Zhao, X. Yuan, and S.C. Benjamin,
\textit{High-Threshold Code for Modular Hardware With Asymmetric Noise},
Phys. Rev. Appl. \textbf{12}, 064006 (2019).

\bibitem{Tuckett2020} D. K. Tuckett, S. D. Bartlett, S. T. Flammia, and B. J. Brown,
\textit{Fault-tolerant thresholds for the surface code in excess of 5\% under biased noise},
Phys. Rev. Lett. \textbf{120}, 050505 (2018).



\bibitem{Bennett1996} C. H. Bennett, D. P. DiVincenzo, J. A. Smolin, and W. K. Wootters,
\textit{Mixed state entanglement and quantum error correction},
Phys. Rev. A \textbf{54}, 3824 (1996).

\bibitem{Emerson2007} J. Emerson, M. Silva, O. Moussa, C. Ryan, M. Laforest, J. Baugh, D. G. Cory, and R. Laflamme, 
\textit{Symmetrised characterisation of noisy quantum processes}
Science \textbf{317}, 1893(2007)

\bibitem{Dankert2009} C. Dankert, R. Cleve, J. Emerson, and E. Livine,
\textit{Exact and approximate unitary 2-designs and their application to fidelity estimation}
Phys. Rev. A \textbf{80}, 012304 (2009).

\bibitem{Geller2013} M. R. Geller and Z. Zhou,
\textit{Efficient error models for fault-tolerant architectures and the Pauli twirling approximation}
Phys. Rev. A \textbf{88}, 012314(2013).



\bibitem{Gottesman1998} D. Gottesman,
\textit{Theory of fault-tolerant quantum computation},
Phys. Rev. A \textbf{57}, 127 (1998).



\bibitem{Tomita2014} Y. Tomita and K. M. Svore.,
\textit{Low-distance surface codes under realistic quantum noise}
Phys. Rev. A. \textbf{90}, 062320 (2014).

\bibitem{Xu2018} X. Xu, N. D. Beaudrap, J. O’Gorman, and S. C. Benjamin,
\textit{An integrity measure to benchmark quantum error correcting memories}
New J. Phys. \textbf{20}, 023009 (2018).

\bibitem{Chao2018} R. Chao and B. W. Reichardt,
\textit{Quantum error correction with only two extra qubits}
Phys. Rev. Lett. \textbf{121}, 050502 (2018).



\bibitem{DiVincenzo2007} D. P. DiVincenzo and P. Aliferis,
\textit{Effective fault-tolerant quantum computation with slow measurements}
Phys. Rev. Lett. \textbf{98} 020501(2007).


\bibitem{Poulin2006} D. Poulin,
\textit{Optimal and Efficient Decoding of Concatenated Quantum Block Codes}
Phys. Rev. A. \textbf{74}, 052333 (2006).


\bibitem{Stephens2009} A. M. Stephens and Z. W. E. Evans,
\textit{Accuracy threshold for concatenated error detection in one dimension}
Phys. Rev. A. \textbf{80}, 022313 (2009).



\bibitem{Dennis2002} E. Dennis, A. Kitaev, A. Landahl, and J. Preskill.,
\textit{Topological quantum memory}
J. Math. Phys. \textbf{43}, 4452 (2002).



\bibitem{Kolmogorov2009} V. Kolmogorov,
\textit{Blossom V: A new implementation of a minimum cost perfect matching algorithm},
In Mathematical Programming Computation \textbf{1}, 43 (2009).



\bibitem{Li2016} Y. Li,
\textit{Noise threshold and resource cost of fault-tolerant quantum computing with Majorana fermions in hybrid systems}
Phys. Rev. Lett. \textbf{117}, 120403 (2016).

\bibitem{Puri2020} S. Puri, L. St-Jean, J.A. Gross, A. Grimm, N.E. Frattini, P.S. Iyer, A. Krishna, S. Touzard, L. Jiang, A. Blais, S.T. Flammia, and S.M. Girvin.
\textit{Bias-preserving gates with stabilized cat qubits}
Science Advances. \textbf{6}, eaay5901(2020).


\bibitem{Guillaud2019} J. Guillaud and M. Mirrahimi
\textit{Repetition cat-qubits for fault-tolerant quantum computation with highly reduced overhead}
Phys. Rev. X. \textbf{9}, 041053(2019).

\end{thebibliography}
\end{document}